\documentclass{sig-alternate} 
\usepackage{stmaryrd}

\begin{document}

\title{Analyzing Alloy Constraints using an SMT Solver:\\ A Case Study}
\subtitle{}

\numberofauthors{2}
\author{
\alignauthor
Aboubakr Achraf El Ghazi \\
       \affaddr{Karlsruhe Institute of Technology}\\
       \affaddr{Karlsruhe, Germany}\\
       \email{elghazi@kit.edu}
\alignauthor
Mana Taghdiri\\
       \affaddr{Karlsruhe Institute of Technology}\\
       \affaddr{Karlsruhe, Germany}\\
       \email{mana.taghdiri@kit.edu}
}

\maketitle

\begin{abstract}
This paper describes how Yices, a modern SAT Modulo theories solver, can be used to analyze the address-book problem
expressed in Alloy, a first-order relational logic with transitive closure. Current analysis of Alloy models -- as performed by
the Alloy Analyzer -- is based on SAT solving and thus, is done only with respect
to finitized types. Our analysis generalizes this approach by taking advantage of the background theories available
in Yices, and avoiding type finitization when possible. Consequently, it is potentially capable of proving that an assertion is a
tautology -- a capability completely missing from the Alloy Analyzer. This paper also reports on our experimental results that
compare the performance of our analysis to that of the Alloy Analyzer for various versions of the address book problem.
\end{abstract}





\section{Introduction}

Alloy\cite{alloy-book} is a first order, declarative language that is widely used for specifying safety properties of 
structurally-rich systems. It is based on relational logic and supports transitive closure as a built-in language construct.
Due to its expressiveness and yet simplicity, Alloy has been used in a wide range of applications, both as a stand-alone
constraint solver (e.g \cite{mondex-alloy, alloy-filesystem, alloy-network}), and as
a backend engine in various program analysis tools (e.g \cite{jalloy, forge, taghdiri-ase-journal, testera}).

Alloy models can be analyzed fully automatically. However, the analysis is always performed with respect to a bounded
{\em scope} in which only a finite number of values is considered for each type. This is because the constraints are translated to
a propositional logic and solved using a SAT solver. Therefore, although the Alloy Analyzer can produce
counterexamples efficiently, it can never {\em prove} the correctness of an assertion -- even for the simplest models. Furthermore,
since arithmetic expressions are directly translated to SAT via bit blasting, they can be analyzed with respect to only a few bits.
Consequently, Alloy offers limited support for arrays and numerical constraints. 

This motivated our project: to analyze Alloy models using an SMT solver rather than a SAT solver. SMT solvers are particularly
attractive because they can efficiently prove a rich combination of decidable background theories without sacrificing completeness or full automation.  
Furthermore, their capability to generate satisfying instances as well as unsatisfiable cores\cite{unsat-core} 
(offered only by some SMT solvers) supports Alloy's lightweight and easy-to-use approach. 

This paper describes the first step of our project. It reports on a case study where an SMT solver, namely Yices\cite{yices-website},
is used to analyze an Alloy model, namely the address book problem\cite{address-book}. To our knowledge, this is the first attempt to
analyze a rich relational logic using an SMT solver. 

We have checked several 
assertions in three different versions of the address book model: (1) the basic model where each name is mapped to at most one
address, (2) the hierarchical model where groups and aliases are allowed, and (3) the acyclic model where no name is 
mapped to itself. Although the models are small, their constraints are typical of Alloy formulas; they
include many of the Alloy constructs that are often used in various applications. 

This case study is performed in the context of a bigger project in which Alloy specifications will be automatically translated 
to an SMT logic and solved by an SMT solver. Therefore, we ensure that
our formulation of the address-book problem is loyal to its Alloy model. However, in order to mitigate the 
bounded-analysis problem of Alloy, we avoid type finitization as much as possible. This approach poses challenges in handling 
some Alloy constructs such as abstract types, multiplicity keywords on a relation's range, and  set membership.
Our translation of these constructs involves the use of Yices $\lambda$-expressions and quantifiers.

Since the Alloy logic is undecidable, type finitization is inevitable for some Alloy constructs. In particular, our
encoding of transitive closure requires type finitzation. However, even when finitization is required, it can be done
on-demand: it is sufficient to finitize 
only the types to which those certain constructs are applied; the other types can stay unbounded and be interpreted as infinite. 
Therefore, checking Alloy models using an SMT solver provides a more general analysis than using a SAT solver.

Our experimental results are very encouraging. Out of a total of 9 assertions 
checked, 5 could be verified by Yices without having to finitize any of the types, meaning that Yices could soundly prove them,
and the other 4 needed only a partial finitization. In most cases, Yices out-performed the Alloy Analyzer in terms of the analysis time.

The rest of the paper is organized as follows: 
 Section \ref{overview} provides an overview of the approach along with some background on Alloy and Yices. Section \ref{case-study}
 gives the details of our case study. Section \ref{evaluation} reports on the performance results. Section \ref{related} describes related
 work and Section \ref{conclusions} concludes the paper. 
 

\section{Overview} \label{overview}

\subsection{Background on Alloy} \label{alloy}

Alloy is a first order relational logic with an object-oriented-like syntax. An Alloy model consists of a type
declaration part, a number of formulas (facts), and an assertion. The Alloy Analyzer checks the validity of the assertion
with respect to a user-provided {\em scope}, an upper bound on the number of elements considered for each type. 
In case the assertion is not valid, the analyzer produces a counterexample with symbolic values for each type and relation.
In this section, we briefly describe a subset of the Alloy language used in our case study. More details can be
found elsewhere\cite{alloy-book}.  

{\bf Types.}
Alloy types represent sets of atoms and are introduced using the signature construct. The declaration $sig\ A \{\}$ introduces a 
top-level type named $A$. A type can also be introduced as a subtype (subset) of another type using either the $extends$ or the $in$ keyword. 
The declarations $sig\ B\ extends\ A \{\}$ and $sig\ C\ extends\ A \{\}$ define $B$ and $C$ as two disjoint subsets of $A$, whereas the 
declarations $sig\ B\ in\ A \{\}$ and $sig\ C\ in\ A \{\}$ allow $B$ and $C$ to have common elements. The $abstract$ keyword preceding
a signature $A$ constrains all atoms of type $A$ to belong to one of its subtypes.

A signature declaration may also contain fields (relations). The declaration $sig\ A\ \{ f:\ B\}$ declares $A$ as a top-level type
and $f$ as a relation of type $A \rightarrow B$. By default, $f$ will be a {\em total function}. That is, each element of $A$ will be mapped to
exactly one element of $B$. However, the multiplicity can be changed using the keywords $lone$ (at most one), $some$ (at least one),
$one$ (exactly one), and $set$ (any number). 

{\bf Expressions.}
Every Alloy expression is a relation. The number of columns in a relation is called its {\em arity} and 
the number of rows is called its {\em size}.  Sets are unary relations, and scalars are singleton unary relations.

The standard set operators union, intersection, and difference are
denoted by "+" , "\&", and "-" respectively. The "." operator denotes relational join: for two relations $p$ and $q$ with arities
$m$ and $n$ respectively, the expression $p.q$ is a relation of arity $m+n-2$ defined as 
$\{ (p_1, \dots, p_{m-1}, q_2, \dots, q_n)\ |\\ \ (p_1, \dots, p_{m}) \in p\ \wedge\ (q_1, \dots, q_n) \in q\ \wedge\ p_m = q_1\}$. 
The $p \rightarrow q$ expression denotes Cartesian product of the two relations $p$ and $q$.

The operators \verb|^| and \verb|*| respectively denote transitive closure and reflexive transitive closure, and are defined only 
on homogeneous binary relations.

{\bf Formulas.}
Basic Alloy formulas are formed by the use of equality operator "=" and the subset operator "in". The formula $p : q$ is the same
as $p\ in\ q$ with the additional constraint that if $q$ is a set, then $p$ should be a scalar. 

Quantified formulas have the general form of $Q\ x: e\ |\ F$
where $F$ is a formula based on the variable $x$, the expression $e$ bounds the values of $x$, and $Q$ is a quantifier.
In addition to the standard universal ($all$) and existential ($some$) quantifiers, Alloy also offers $one$ (exactly one),
$lone$ (at most one), and $no$ (none) quantifiers. All quantifiers except $all$ can also be used with an expression alone to
constrain its cardinality. The formula $lone\ A$, for example, constrains the relation $A$ to have at most one element.

{\bf Facts, Predicates, Functions and Assertions.}
Non-parametrized constraints (assumptions) of the system are expressed as {\em facts}. These constraints are
considered to be true all the time. Parameterized or reusable formulas 
expected to be used in different contexts are expressed as {\em predicates} and reusable expressions are expressed as 
{\em functions}. The properties to be checked are expressed as {\em assertions}. 

\subsection{Background on Yices}

Yices is a modern SAT Modulo Theories (SMT) solver that checks satisfiability of arbitrary formulas containing 
uninterpreted function symbols with equality, linear real and integer arithmetic, scalar types, recursive datatypes, 
dependent types, tuples, records, extensional arrays, fixed-size bit-vectors, quantifiers, and $\lambda$-expression \cite{yices}. 
It can also compute MAX-SMT and produce unsatisfiable cores.
Yices accepts the SMT-Lib format\cite{smt-lib} as input. However, it also supports a LISP-like language that is
more expressive than SMT-Lib. This is the language that we use in our case study.  In this section, we describe a subset
of the Yices language that we use. More details can be found elsewhere\cite{yices-website}.

{\bf Types and Subtypes.}
In addition to its built-in types \texttt{real, int, nat}, and \texttt{bool}, Yices allows users to declare new basic (uninterpreted) types.
The type declaration \texttt{(define-type A)} defines $A$ as a new uninterpreted type, whereas \texttt{(define-type A e)}
defines $A$ as an alias for the type expression $e$ which is expressed in terms of the existing types. 

A type expression \texttt{(-> $A_1 \dots A_n$)} denotes a function type over the types $A_1$ to $A_n$. 
The expression \texttt{(scalar $a_1 \dots a_n$)} denotes a scalar type consisting of the identifiers $a_1$ to $a_n$. 
The type expression \texttt{(subtype ($x$::$A$) $p$)} denotes a subtype of $A$ for which the predicate
$p$ holds. 

{\bf Expressions.}
A constant value, function, or predicate $x$ of type $T$ can be declared as \texttt{(define x::T)}. A constant
can also be defined as a particular expression $e$ using \texttt{(define x::T e)}.

The usual boolean operators \texttt{and}, \texttt{or}, \texttt{not}, and \texttt{=>} (implies) are allowed.
Equality and inequality are  denoted by \texttt{=} and \texttt{/=} respectively.
Conditional expressions can be expressed using the if-then-else semantics denoted by the ternary
operator \texttt{(if cond $e_1$ $e_2$)}.

$\lambda$-expressions are also allowed. They are generally used to express unnamed functions, 
and have the following syntax:
\texttt{(lambda ($t_1$::$T_1 \dots t_n$::$T_n$) e)}

Functions (among other types) can be updated using the \texttt{update} construct. The expression
\texttt{(update f ($p_1 \dots p_n$) v)} updates the function $f$ at the location $[p_1, \dots, p_n]$ with the
new value $v$. It is semantically equivalent to the $\lambda$-expression:
\texttt{(lambda ($t_1$::$T_1 \dots t_n$::$T_n$) (if (and (= $t_1$ $p_1$) $\dots$ (= $t_n$ $p_n$)) v (f $t_1 \dots t_n$)))}

There is also limited support for recursive functions. Yices expands recursion during pre-processing.
The default recursion limit is 30, but it can be changed using \texttt{(set-nested-rec-limit! n)} where $n$
is a constant number. 

{\bf Commands.}
In addition to the above constructs, the Yices language provides a set of commands including
the  \texttt{(assert f)} command that asserts a formula $f$ in the current logical context.

\subsection{Approach}

We apply the following rules to translate the Alloy address-book model to the Yices language.
It should be noted that these rules are not meant to be complete; they do not address all Alloy 
constructs. They only address so much of the Alloy language that is necessary to translate the
address book model. Here we only provide an overview of the translation; the details are presented
in Section \ref{case-study}.

\begin{itemize}

\item{Predicates, functions.}
All predicates and functions of the Alloy model are inlined at their usage sites.

\item{Facts.}
An Alloy fact $f$ is translated to $\llbracket f\rrbracket$ and added to the Yices constraints
using the command $(assert\ \llbracket f\rrbracket)$

\item{Assertions.} 
An Alloy assertion $f$ is negated, translated, and added to the Yices constraints as $(assert\ \llbracket \neg f\rrbracket)$.
Therefore, if Yices finds a satisfying instance, that instance will represent a counterexample to the assertion.
Otherwise the assertion is valid (with respect to the finite bounds of types if any).

\item{Types.} 
A basic signature $A$ in Alloy is translated to a basic, uninterpreted type in Yices.
An Alloy extension type $sig\ B\ extends\ A\{\}$ becomes a Yices subtype \texttt{(define-type B (subtype (a::A) (isB a)))}
where \newline 
the \texttt{isB} function (of type \texttt{A -> bool}) determines which elements of $A$ are instances of $B$. 
Additional axioms are used to ensure that multiple extension types are disjoint,
and to enforce the semantics of abstract signatures (see Section \ref{case-study}).

\item{Relations.}
An Alloy functional relation $f: A \rightarrow B$ is translated to a Yices function \texttt{(define f::(-> A B))}. 
If $f$ is a partial function, a special constant \texttt{noB} is defined to represent the empty value: \texttt{(define noB::B)}.
A non-functional relation $r: A \rightarrow B$ in Alloy is translated to a Yices function with an additional
boolean column whose value is "true" for the tuples that belong to $r$ and "false" for all others: 
\texttt{(define r::(-> A (-> B bool)))}  

\item{Relational operators.}
The operators union, intersection, relational join, and transitive closure are defined separately for the functional
relations and non-functional relations. Transitive closure is defined recursively and thus, requires type
finitization. Details of these operators are shown in Section \ref{case-study} as needed. 

\item{Quantifiers.}
Universal quantifiers are encoded using type finitization\footnote{We could use Yices quantifiers directly. 
But, in this case study, the only universally quantified formula contains transitive closure. So finitization is more appropriate.}. 
A formula $all\ x:T\ |\ f(x)$ is translated to $(assert\ (\llbracket f\rrbracket\ T_1)) \dots (assert\ (\llbracket f\rrbracket\ T_n))$
where $T_1$ to $T_n$ denote the possible values of type $T$.
Existential quantifiers are skolemized.
 
\end{itemize}


\section{The Address Book Case Study} \label{case-study}
The address-book problem\cite{address-book} models the address book system of  an email client. It represents a 
database that associates email addresses with names. We describe three versions of the address book: (1) the basic
one in which each name is mapped to at most one address, (2) the hierarchical one in which
an {\em alias} name can be created for an address and addresses can be referred to by {\em group} names, and (3)
an acyclic one which is similar to the hierarchical one except for the additional constraint that 
no name can appear in its own set of aliases and groups.

\subsection{Basic Address Book}  \label{basic}
 
\begin{figure}
{\small
\begin{verbatim}
1:  sig Name, Addr {}
2:  sig Book { 
3:    addr: Name -> lone Addr 
    }
    pred add (b, b': Book, n: Name, a: Addr) {
4:    b'.addr = b.addr + n->a
    }
    pred del (b, b': Book, n: Name) {
5:    b'.addr = b.addr - n->Addr
    }
    fun lookup(b: Book, n: Name): set Addr {
6:    n.(b.addr)
    }
    assert delUndoesAdd{
7:    all b, b', b'': Book , n: Name, a: Addr |
8:      no n.(b.addr)  and 
9:      add[b, b', n, a] and 
10:     del[b', b'', n] implies 
11:       b.addr = b''.addr
    }
12: check delUndoesAdd for 3
\end{verbatim}
}
\caption{Basic address book model in Alloy}
\label{list:basic_address_book_alloy}
\vspace*{-0.3cm}
\end{figure}

The basic address book model is given in Figure \ref{list:basic_address_book_alloy}.
Lines 1-3 declare three basic types $Name$, $Addr$, and $Book$, and a ternary functional relation  $addr:\ Book \rightarrow Name \rightarrow Addr$ that maps 
each pair $(b, n)$ of $Book$ and $Name$ to at most one $Addr$.

To describe the dynamic behavior of the system, the model defines two predicates: {\em add} 
for the addition operation and {\em del} for deletion. The lookup function returns all the addresses that correspond to a name in a particular book.

The original model contains three assertions that check how different combinations of these operations behave. 
In the interest of space, we only discuss the {\em delUndoesAdd} assertion. As the name suggests, this assertion specifies that if a fresh name and an address
are first added to a book, and then deleted, the resulting book is the same as the original one.  
The assertion holds and thus, no counterexamples can be found.

The Alloy Analyzer, however, cannot prove that this assertion is a tautology. It can only check the model with respect to a bounded scope
given by the user. In this case, a scope of 3 is provided in Line 12.

\begin{figure}
{\small
\begin{verbatim}
1: (define-type Name) 
1: (define-type Addr)  
2: (define-type Book)
  
3: (define addr::(-> Book (-> Name Addr)))
3: (define noAddr::Addr)

7: (define b::Book)
7: (define b'::Book)
7: (define b''::Book)
7: (define n::Name)
7: (define a::Addr)
7: (assert (/= a noAddr))   

8: (assert (= ((addr b) n) noAddr))   
9: (assert (or (= ((addr b) n) a) (= ((addr b) n) noAddr)))
9: (assert (= (addr b') (update (addr b) (n) a)))       
10:(assert (= (addr b'') (update (addr b') (n) noAddr)))  
11:(assert (/= (addr b) (addr b'')))    
\end{verbatim}
}
\caption{Translation of basic adress-book to Yices}
\label{list:basic_address_book_yices_assertion01}               
\end{figure}

Figure \ref{list:basic_address_book_yices_assertion01} gives our translation of the basic address book to Yices.
The numbers in this figure denote which line in the Alloy model has produced a particular Yices constraint.
The translation steps are described below:

\begin{itemize}
 \item (Lines 1-2). The Alloy basic types $Name$, $Addr$, and $Book$ are translated to uninterpreted types in Yices.
 
 \item (Line 3). The relation $addr$ is translated to a ternary function that maps each Book to a function from Name to Addr. Since 
 Yices functions map each element of the domain type to exactly one element of the range type, we translate the {\em lone}
 multiplicity construct by introducing a special Yices constant $noAddr$. This constant represents a non-value of type
 $Addr$. That is, if $((addr\ b)\ n) = noAddr$, then the name $n$ is not mapped to any addresses in the book $b$.

\item (Lines 7-11). In order to find a counterexample for an assertion, we add its negation to the set of Yices constraints.
The negation of the assertion {\em delUndoesAdd} is 
{\small
\begin{verbatim}
7: some b, b', b'': Book , n: Name, a: Addr |
8:     no n.(b.addr)  and 
9:     add[b, b', n, a] and 
10:    del[b', b'', n] and 
11:    not (b.addr = b''.addr)
\end{verbatim}
}
and its translation is given in Lines 7-11 of Figure \ref{list:basic_address_book_yices_assertion01}. 
The constant definitions (Line 7) correspond to the existential quantifier in Alloy. Since a non-value has been
defined for the $Addr$ type in Yices, in order to follow the semantics of
$a: Addr$ in Alloy, we constrain the constant $a::Addr$ not to be non-value ($noAddr$).

The translation of Line 8 exploits the semantics of the $noAddr$ constant.

The union operator used in the {\em add} operation (Line 9) can be efficiently translated using Yices function updates.
For a functional relation $f: X \rightarrow Y$, the Alloy expression $f + x \rightarrow y$ can be expressed by the Yices
expression $(update\ f\ (x)\ y)$. However, because $f$ is functional, if it already contains a pair $(x, z)$ where $z \ne y$,
then the above union operation is undefined. Therefore, in order to follow the Alloy semantics, our Yices translation of this
union expression constrains  $f(x)$ to be either empty or equal to $y$.
 
The {\em del} operation (Line 10) removes all mappings of a name $n$ from the book $b'$. This can be translated using
an update that maps $n$ to the non-value $noAddr$. 

The last line of the Yices model (Line 11) is a straightforward translation of the inequality constraint in the negation of the Alloy assertion. 
\end{itemize}

When checking {\em delUndoesAdd}, Yices outputs "unsat", meaning that no counterexample exists.
Since the constraints are checked for infinite types, the unsat result is a {\em proof of correctness}. That is, 
unlike Alloy, Yices can show that this assertion is a tautology.

\subsection{Hierarchical Address Book}

The hierarchical model represents a more realistic address book. It allows to create an alias for an address and then use that 
as the target address of another alias. It also allows to use an alias for multiple targets so that a group of addresses can 
be referred to by a single name. The hierarchical address book is given in Figure \ref{list:hierarchical_address_book_model_alloy}.

\begin{figure}
{\small
\begin{verbatim}
1:  abstract sig Target {}
2:  sig Addr extends Target {}
3:  abstract sig Name extends Target {}
4:  sig Alias, Group extends Name {}
5:  sig Book {
6:    names: set Name,
7:    addr: names -> some Target
     }
    fact {
8:    all b: Book, a:Alias | lone a.(b.addr)
    }
    pred add (b, b': Book, n: Name, t: Target) {
9:    b'.addr = b.addr + n->t
    }
    pred del (b, b': Book, n: Name, t: Target) {
10:   b'.addr = b.addr - n->t
    }
    fun lookup (b: Book, n: Name): set Addr {
11:   n.^(b.addr) & Addr
    }
    assert delUndoesAdd {
12:   all b, b', b'': Book , n: Name, t: Target |
13:     no n.(b.addr) and 
14:     add[b, b', n, t] and 
15:     del[b', b'', n, t] implies 
16:        b.addr = b''.addr 
    }
\end{verbatim}
}
\caption{Hierarchical address book model in Alloy} 
\label{list:hierarchical_address_book_model_alloy}
\vspace*{-0.3cm}
\end{figure}

The major differences between the hierarchical model and the basic one are the use of the Alloy type hierarchy and multiplicity constructs.
The operations and assertions are very similar to the ones in the basic model\footnote{The lookup function here uses the transitive closure
operator which will be discussed in Section 3.3.}. The corresponding Yices translation is given in Figure \ref{list:hierarchical_address_book_yices}.
Again, the line numbers in this figure denote which lines in the Alloy model have produced which Yices constraints.
The main ideas of this translation are described below.

\begin{figure}
{\small
\begin{verbatim}
1:  (define-type Target)

2:  (define isAddr::(-> Target bool)) 
2:  (define-type Addr (subtype (t::Target) (isAddr t)))

3:  (define isName::(-> Target bool)) 
3:  (define-type Name (subtype (t::Target) (isName t)))

3:  (assert (forall (t::Target) 
        (not (and (isAddr t) (isName t)))))
3:  (assert (forall (t::Target) (or (isAddr t) (isName t))))

4:  (define isAlias::(-> Name bool)) 
4:  (define-type Alias (subtype (n::Name) (isAlias n)))

4:  (define isGroup::(-> Name bool)) 
4:  (define-type Group (subtype (n::Name) (isGroup n)))

4:  (assert (forall (n::Name) 
        (not (and (isAlias n) (isGroup n))))) 
4:  (assert (forall (n::Name) (or (isAlias n) (isGroup n))))

5:  (define-type Book)

6:  (define names::(-> Book (-> Name bool)))

7:  (define-type addrRange (-> Name (-> Target bool)))
7:  (define-type addrType (-> Book addrRange))
7:  (define choose::addrType)
7:  (define oneTarget::(-> Book (-> Name Target)))

7:  (define addr::addrType 
7:      (lambda (b::Book)
7:          (lambda (n::Name)
7:              (lambda (t::Target)
7:                  (if (not ((names b) n)) 
7:                       false
7:                       (if (= t ((oneTarget b) n))
7:                            true
8:                            (if (isAlias n)
8:                                 false
7:                                 (((choose b) n) t)
    )))))))

12: (define b::Book)                   
12: (define b'::Book)                   
12: (define b''::Book)                  
12: (define n::Name)                    
12: (define t::Target)                  

    (define f::addrRange)
    (assert (= f (addr b)))
    (define f'::addrRange)
    (assert (= f' (addr b')))
    (define f''::addrRange)
    (assert (= f'' (addr b'')))
      
13: (define emptyTarget::(-> Target bool) 
        (lambda (t::Target) false))
13: (assert (= (f n) emptyTarget))                                          
14: (assert (= f' (update f (n) (update (f n) (t) true))))   
15: (assert (= f'' 
        (update f' (n) (update (f' n) (t) false))))    
16: (assert (/= f f''))                                                           
\end{verbatim}
}
\caption{Translation of hierarchical address book to Yices} 
\label{list:hierarchical_address_book_yices}
\end{figure}

\begin{itemize}
 \item (Lines 1-5). The type hierarchy of the Alloy model is translated using the Yices {\em subtype} construct along with uninterpreted membership functions. 
The extensions of an abstract Alloy signature divide the space of all atoms into disjoint subsets. To avoid finitizing types, we use explicit axioms to constrain the
membership functions accordingly. Such axioms are applied to all levels of the type hierarchy.
 
 \item (Line 6). The Alloy relation $names: Book \rightarrow set\ Name$ is a non-functional relation. Therefore, it is translated to the Yices function
 $(\rightarrow Book\ (\rightarrow Name\ bool))$ in which
 the extra boolean column denotes whether a pair $(b, n)$ belongs to the relation $names$ or not.
 
 \item (Line 7). Similar to $names$, the non-functional relation $addr$ is declared using an additional 
 boolean column, i.e. of type $(\rightarrow Book\ (\rightarrow Name\ (\rightarrow Target\ bool)))$. 
 However, in the Alloy model, for every book $b$, $(addr\ b)$ is defined only for those atoms of type $Name$ that belong to $(names\ b)$.
 We use a $\lambda$-expression to express this fact. The lambda expression specifies that for any book $b$ and name $n$,
 if $((names\ b) n)$ is false, then $(((addr\ b) n) t)$ is also false (for all target $t$).
 
 Furthermore, the {\em  some} keyword in the declaration of $addr$ specifies that any name in the domain set of
 $addr$ is mapped to at least one target. We represent this in Yices using the auxiliary functions $choose$ 
 and $oneTarget$.  The former models the fact that $((addr\ b) n)$ can be a {\em set} of targets, whereas the
 latter models the fact that this set is {\em non-empty}. More precisely, $choose$
 is an unconstrained function that may contain any number of tuples.
The function $oneTarget$, on the other hand, maps every pair $(b, n)$ to exactly one target. 
The $addr$ relation contains all tuples $(b, n, t)$ that belong either to $oneTarget$ or to $choose$
(assuming that $((names\ b) n) = true$).
 
 \item (Line 8) The $lone$ keyword, in the Alloy model, specifies that $addr$ maps each alias to at most one target. 
We augment the Yices $\lambda$-expression defining $addr$ to specify this fact. 
A target is of type $Alias$ if it passes the $isAlias$ test. The $addr$ relation maps an alias
to exactly one target: the one specified by the $oneTarget$ function.
 
 \item (Lines 12-16). The translation of the assertion is similar to the basic address book. The only 
 differences are because of the additional boolean column in the declaration of the $addr$ function.
 The union operator requires a double update of $addr$ and the constraint $no\ n.(b.addr)$ in line 
 13 requires the definition of the auxiliary $emptyTarget$ function. 
 
\end{itemize}

Again, Yices outputs "unsat". This means that the assertion has been proven valid without having
to finitize any of the types -- a result that can never be achieved by the Alloy Analyzer.


\subsection{Acyclic Address Book} \label{acyclic}

The acyclic address book is the same as the hierarchical one except for an extra fact
that states that for any book, there is no name that belongs to the set of targets reachable from the name itself.
That is, $b.addr$ is acyclic.
{\small
\begin{verbatim}
 all b: Book, n: Name | not (n in n.^(b.addr))
\end{verbatim}
}

The challenge of this constraint is to translate transitive closure properly. For a homogeneous relation $r: A \rightarrow A$,
we have $\verb|^|r = r + r.r + r.r.r + \dots + r^{(i)} + \dots$ where the computation of $r^{(i)}$ continues until a fixpoint is reached.
Our translation of transitive closure to the Yices language requires finitzation of the type $A$, and 
is based on the auxiliary functions union, join, and iterative-join defined
for non-functional relations. 

{\bf Union.} The operation $(union\ f\ g)$ returns all the tuples that are either
in $f$ or in $g$. The formal definition is as follows: 
{\small
\begin{verbatim}
(define-type relType (-> A (-> B bool)))

(define union::(-> relType relType relType)
  (lambda (f::relType g::relType)
    (lambda (a::A) 
      (lambda (b::B) 
        (or ((f a) b) ((g a) b))
))))
\end{verbatim}
}

{\bf Join:} The definition of the Alloy join operator is given in Section \ref{alloy}. The operation $(join\ f\ g)$ contains a 
tuple $(a, c)$ if $\exists b\ |\ (f\ a\ b) \wedge (g\ b\ c)$. That is, 
{\small
\begin{verbatim}
(define-type relType1::(-> A (-> B bool)))
(define-type relType2::(-> B (-> C bool)))
(define-type relType3::(-> A (-> C bool)))

(define join::(-> relType1 relType2 relType3)
  (lambda (f::relType1 g::relType2)
    (lambda (a::A)
      (lambda (c::C)
        (exists (b::B) (and ((f a) b) ((g b) c)))
))))
\end{verbatim}
}

{\bf Transitive closure:}
We define a stepwise transitive closure recursively using the iterative-join operator.
For a natural number $i > 0$ and a homogeneous relation $r$, we define $(iterJoin\ i\ r) = r^{(i)}$ and
the transitive closure $(tc\ i\ r) = r + r.r + \dots + r^{(i)}$ recursively.
{\small
\begin{verbatim}
(define-type relType::(-> A (-> A bool)))

(define iterJoin::(-> nat relType relType)
  (lambda (i::nat r::relType)
    (if (= i 1) r (join r (iterJoin (- i 1) r)))
))

(define tc::(-> nat relType relType)
  (lambda (i::nat r::relType)
    (if (= i 1) r (union (tc (- i 1) r) (iterJoin i r)))
))
\end{verbatim}
}

It is easy to see that if the type $A$ consists of only $n$ distinct values, then $\verb|^|r = (tc\ n\ r)$. That is, 
it is guaranteed that after at most $n$ steps, $\verb|^|r$ reaches a fixpoint. 

Having defined a transitive-closure operator, we translate the acyclicity constraint by finitizing the types
$Book$ and $Name$ to $n$ values, inlining the universal quantifiers for all those
values, and replacing the transitive closure operator with $(tc\ n)$. 

In finitized models,
the "unsat" ouput of Yices only means that the assertion holds with respect to the analyzed finitization bounds.
No general proof of validity is implied. 
However, because in the Yices model, the types $Target$ and $Addr$ are not finitized, the analysis accounts for
a larger scope, and thus, the outcome is more general than Alloy's outcome.
The results of checking various assertions in the acyclic model with different bounds for types are discussed in the
next section.


\section{Evaluation}  \label{evaluation}

\begin{table*}
\centering
 \begin{tabular}[c]{|c|c|c|c|c|c|}
\hline
 Model & Assertion & Scope & Yices time (s) & Alloy time (s)& Tautology?\\
\hline
%
%
Basic Addr. Book     	&  delUndoesAdd  & 20 & 0.0006 & 4.36 		& Yes \\
			&                & 25 & 0.0006 & 13.11 		&     \\ 
			&                & 30 & 0.0006 & 45.82 		&     \\ 
			&                & 35 & 0.0006 & time-out 	&     \\ 
\cline{2-6}
			&  addIdempotent 
			                & 30 & 0.0006 & 82.38 		&  Yes   \\ 
			&                & 40 & 0.0006 & 141.27 	&     \\ 
			&                & 50 & 0.0006 & time-out 	&     \\ 
\cline{2-6}
			&   addLocal   & 40 & 0.0003 & 19.67 		&  Yes   \\
			&                & 50 & 0.0003 & 45.01 		&     \\ 
			&                & 60 & 0.0003 & 102.41 	&     \\ 
			&                & 70 & 0.0003 & memory-out 	&     \\ 
\hline
%
%
Hierarchical Addr. Book & delUndoesAdd   & 20 & 0.009  & 5.44           & Yes \\
                         &               & 30 & 0.009  & 54.02 	 	&     \\
                        &                & 40 & 0.009  & 139.25 	&     \\ 
                        &                & 50 & 0.009  & time-out 	&     \\ 
\cline{2-6}
                        & addIdempotent  & 20 & 0.008  & 7.46     	& Yes \\ 
                        &                & 30 & 0.008  & 23.71    	&     \\ 
                        &                & 40 & 0.008  & time-out 	&     \\ 
\cline{2-6}
                        & addLocal       & $n$ = 2 & 0.02     & 0.19 	&  No \\ 
                        &                & $n$ = 3 & time-out     & 0.13 	&     \\ 
\hline
%
%
Acyclic Addr. Book      &  delUndoesAdd  
                                        & $n$ = 4 & 0.45     & 0.19     &  Don't know           \\ 
                        &                & $n$ = 5 & 14.36    & 4.44     &            \\ 
                        &                & $n$ = 6 & 8.10     & 150.31   &            \\ 
                        &                & $n$ = 7 & 29.45    & time-out &            \\ 
\cline{2-6}
                        & addIdempotent  
                                        & $n$ = 4 & 0.39     & 0.22     &  Don't know          \\ 
                        &                & $n$ = 5 & 10.76    & 4.03     &            \\ 
                        &                & $n$ = 6 & 8.10     & 135.59   &            \\ 
                        &                & $n$ = 7 & 30.20    & time-out &            \\ 
\cline{2-6}
                        & addLocal       & $n$ = 2 & 0.07     & 0.18     & No         \\ 
                        &                & $n$ = 3 & time-out & 0.23     &            \\
\hline
 \end{tabular}
 \caption{Performance evaluation results}
 \label{tab:measurement:basic_address_book}
 \vspace*{-0.5cm}
\end{table*}

We have evaluated our translation of Alloy to Yices by checking the 3 assertions of the Alloy address book model.
In addition to the {\em delUndoesAdd} assertion discussed before, we check the two assertions {\em addIdempotent}
and {\em addLocal} given in Figure \ref{assertions}. The first one states that repeating an addition has no effect,
and the second one states that adding an entry for a name $n$ does not affect the result of a lookup for a different name $n'$.
All assertions are checked in all 3 versions of the address book. The assertions delUndoesAdd and addIdempotent have
no counterexamples in any of the models. The addLocal assertion, however, is valid only in the basic version. In the other two address books,
it has a counterexample.

\begin{figure}
{\small
\begin{verbatim}
assert addIdempotent {
  all b, b', b'': Book, n: Name, t: Target |
    add[b, b', n, t] and add[b', b'', n, t] implies 
      b'.addr = b''.addr
}

assert addLocal {
  all b, b': Book, n, n': Name, t: Target |
    n!=n' and add[b, b', n, t] implies 
      lookup[b,n'] = lookup[b', n']
}
\end{verbatim}
}
\caption{Other assertions of the Alloy model}
\label{assertions}
\end{figure}

We evaluated the correctness of our translation by ensuring that whenever the Alloy Analyzer returns a counterexample, Yices 
returns a valid counterexample too, and when Alloy cannot find a counterexample in a specific scope, Yices does not find any 
in that scope either. We evaluated the efficiency of our translation by comparing the Yices analysis time to that of Alloy.
The results are given in Table \ref{tab:measurement:basic_address_book}.
The {\em time} columns give the CPU time (in second) measured on an Intel Core 2 Quad CPU 2.83GHz with 8GB memory. 
The time-out threshold is 180 seconds. We increase the analysis scope until
either Alloy or Yices times out. The Alloy analysis time is the total of the time spent on generating CNF and the time spent in the SAT 
solver as reported by the Alloy Analyzer 4.1.10 running the SAT4J solver. The Yices analysis time is what Yices 1.0.27 reports using the {\em -st} option.
We have repeated each experiment 3 times and given the average analysis time in the tables. 

The {\em Tautology?} column
is "Yes" if Yices manages to prove the correctness of the assertion. That is, if our Yices model does not have any type
finitization and the analysis outputs "unsat" for the negation of the assertion. This column is "No" when a counterexample
is found, and "Don't know" when the Yices model requires finitization of some types. In these cases, even if the analysis returns "unsat",
it cannot guarantee the correctness of the assertion beyond the scope checked.

As shown in Table \ref{tab:measurement:basic_address_book}, in cases where a tautology is proven, our Yices model is very easy to analyze.
Yices runtime for all such assertions is close to zero. Since these cases do not require any type finitization,
increasing the scope has no effects on the Yices analysis time. The Alloy analysis time, however, increases as the scope is increased.
The scope numbers reported in these cases denote the bounds used for all Alloy signatures. For a scope of $x$, we use the Alloy
command \texttt{check [assertion] for x} to perform the analysis. This command causes the Alloy Analyzer to check all configurations
of types and subtypes in which all types have at most $x$ elements. It should be noted that, due to the finitization requirement, 
Alloy can never prove that an assertion is a tautology.  Therefore, in these cases, Yices analysis result is strictly stronger than that of
Alloy.

Checking {\em delUndoesAdd} and {\em addIdempotent} in the acyclic model requires finitizing the types $Book$ and $Name$
(see Section \ref{acyclic}). The value of $n$ given in the scope column of the table gives the bound on these types.
Although our Yices model does not finitize $Target$ and $Addr$, we have to finitize those in Alloy.
To allow at least $n$ atoms of type $Addr$, we use the Alloy command \texttt{check [assertion] for 2n but n Book, n Name}.
For these two assertions, the performance of Yices varies significantly with the scope. As shown in the table, similar to SAT solvers,
the performance of SMT solvers is not monotonic; the Yices analysis time for the scope of 6 is smaller than that of scope 5. Furthermore,
in analyzing finite models containing transitive closure, the Alloy Analyzer can sometimes be more efficient than Yices.

The $addLocal$ assertion (in the hierarchical and the acyclic model) is the only assertion for which Yices performs strictly worse 
than Alloy. Since this assertion uses transitive closure (see the Lookup function of Figure \ref{list:hierarchical_address_book_model_alloy}), 
our Yices model is finitized based on a number $n$ as described above. The assertion is invalid. Similar to the Alloy Analyzer, 
Yices can find a counterexample even in the small scope of $n = 2$. Although this result is sufficient to show that the assertion
is invalid, we increased
the scope to assess the performance of our Yices model. As shown in the table, increasing the scope to 3 makes our model
too difficult for Yices to solve. This is because this assertion
uses the transitive closure operator in both sides of equality. In other words, it asserts that the results of two
transitive closures are equal. Since our definition of transitive closure is a recursive $\lambda$-expression, checking
such equalities becomes too difficult for Yices. Currently, we are investigating other translation techniques for transitive closure
to reduce the complexity of such assertions. 

It should be noted that because of our extensive use of $\lambda$-expressions, when Yices reports a satisfying instance,
it is preceded by the word "unknown". This implies that the instance
might be a false alarm. In our experiments, however, the instances found by Yices were always real counterexamples.
The unsat outputs were always definite, meaning that the assertions were soundly proven correct.


\section{Related Work} \label{related}

SMT solvers have been used as the analysis engine of various software verification tools. To our knowledge, 
however, they have never been used to analyze a relational logic. They are widely used to increase
the automation level of theorem provers, to improve the performance of bounded model checkers, and also as solvers
for specific logics applicable to software verification.

Theorem provers such as PVS\cite{pvs-website}, Key\cite{key-website}, HOL-Light\cite{hol-light-website}, and Isabelle\cite{isabelle-website} 
have integrated SMT solvers in their backend engines to 
improve their automation and support for counterexamples. The \texttt{smt} tactic\cite{smt-tactic}, for example, integrates generic SMT solvers
with Isabelle via translation to SMT-Lib\cite{smt-lib}. The \texttt{ismt} tactic\cite{ismt-tactic}, on the other hand, uses the Yices input language
to take advantage of a wider range of background theories in the translation of Isabelle. Although the use of SMT-Lib as an interface to the
SMT solver allows the use of
different SMT solvers, its limitations can make the translation unintuitive and sometimes less expressive.

SMT solvers are also used for software model checking (e.g. \cite{mcmt, smt-cbmc}). 
SMT-CBMC\cite{smt-cbmc}, for example, integrates CVC-Lite SMT solvers with the C Bounded Model Checker (CBMC). The CBMC tool encodes
program traces in propositional logic and solves them using a pure SAT solver. SMT-CBMC,  on the other hand, provides an encoding into richer,
and yet decidable set of theories supported by CVC-Lite. It produces more compact formulas that are often easier to solve than the ones generated 
by CBMC. 

Botincan, et. al.\cite{botincan} introduced a technique for modular verification of C programs against specifications written in separation
logic. They extended the separation logic prover so that it can use the Z3 SMT solver both as a prover and as a guide for proof search.
This prevents the separation logic prover from stopping at formulas that it cannot reason about. Leino, et. al.\cite{leino} use SMT solvers
within the context of the \verb|Spec#| program verifier. They translate common comprehension expressions into verification conditions that
can be solved by either Simplify or Z3 SMT solvers. A comprehension expression is an expression in which a set of elements
(with a particular characteristics) is combined using an operator (e.g. addition, comparison, multiplication, etc.). Encoding such expressions
in an SMT logic requires the design of appropriate matching triggers. Since the Alloy language also allows set comprehension, we believe
that Leino's approach can be used in our translation of the Alloy language to an SMT formula.


\section{Conclusions} \label{conclusions}

We have described a case study in which the Alloy address book problem is analyzed using the Yices SMT solver.
This case study is the first step of a bigger project in which the Alloy language will be translated to the Yices input
language automatically, and thus, analyzed using an SMT solver rather than a SAT solver. The main advantage of
this approach is to avoid finitization of types when possible, and thus to provide a stronger analysis than the one
offered by the current Alloy Analyzer. 

Since the Alloy Analyzer translates all language constructs to propositional logic and uses
a SAT solver, it performs the analysis only with respect to a finite scope. It can never prove that an assertion is a tautology,
even for the simplest models. Unlike SAT solvers, SMT solvers support a number of theories and can prove or refute
the constraints within those theories without sacrificing completeness.

Our case study was a witness to feasibility of this project. Out of a total of 9 assertions that were checked 
in 3 models, 5 were proven to be tautologies. That is, all Alloy constraints could be translated to the Yices
without having to finitize any type. In the other 4, finitization was required, but only for a subset of the types;
the others were left infinite.  

The Alloy logic is undecidable. Therefore, finitization is inevitable. Our case study, however, showed that
the finitization can be done on-demand; only for those types to which certain language constructs are applied (e.g. transitive closure).

During this case study, we realized that many of the Alloy constructs can be translated to Yices in more than
one way. So far, we have picked the ones that work reasonably well for the address book problem. More experiments
are needed until we can fix a particular translation rule for each Alloy construct. In fact, our current experiments
show that our translation of transitive closure is too difficult to analyze when it is applied to different relations and then
checked for equality. We are currently investigating other translation techniques to mitigate this problem.

Although, in our experiments, any time that Alloy did not find a counterexample, Yices did not either, this is not
always the case. Some Alloy models for which the analyzer cannot find a counterexample, actually have counterexamples,
but in higher scopes than the one that the analyzer can check. Investigating 
whether Yices can analyze such models in a high-enough scope to find the counterexample, will be left for future.

Once the basic translation of the Alloy language is done, we will also investigate how to apply optimization techniques
such as symmetry breaking and subexpression sharing detection to produce formulas that are easier to solve for the underlying SMT solver.



\bibliographystyle{abbrv}
\bibliography{sigproc}  


\end{document}